\documentclass[sigconf, nonacm]{acmart}

\AtBeginDocument{%
  \providecommand\BibTeX{{%
    \normalfont B\kern-0.5em{\scshape i\kern-0.25em b}\kern-0.8em\TeX}}}

\setcopyright{none}
\renewcommand\footnotetextcopyrightpermission[1]{} 
\usepackage{xspace}
\usepackage{mdframed}

\usepackage{lipsum}

\newcommand\blfootnote[1]{%
  \begingroup
  \renewcommand\thefootnote{}\footnote{#1}%
  \addtocounter{footnote}{-1}%
  \endgroup
}

\begin{document}

\title{\rp: A Demo of Retrieval-Enhanced Large Language Models to Support Lateral Reading}

\author{Dake Zhang}
\email{dake.zhang@uwaterloo.ca}
\orcid{0000-0001-9663-9391}
\affiliation{%
  \department{School of Computer Science}
  \institution{University of Waterloo}
  \country{Waterloo, Ontario, Canada}
}
\author{Ronak Pradeep}
\email{ronak.pradeep@uwaterloo.ca}
\orcid{0000-0001-6296-601X}
\affiliation{%
  \department{School of Computer Science}
  \institution{University of Waterloo}
  \country{Waterloo, Ontario, Canada}
}
\renewcommand{\shortauthors}{Zhang and Pradeep}
\newcommand{\rp}{\textsc{ReadProbe}\xspace}

\begin{abstract}

With the rapid growth and spread of online misinformation, people need tools to help them evaluate the credibility and accuracy of online information.
Lateral reading, a strategy that involves cross-referencing information with multiple sources, may be an effective approach to achieving this goal.
In this paper, we present \rp, a tool to support lateral reading, powered by generative large language models from OpenAI and the Bing search engine.
Our tool is able to generate useful questions for lateral reading, scour the web for relevant documents, and generate well-attributed answers to help people better evaluate online information.
We made a web-based application to demonstrate how \rp can  help reduce the risk of being misled by false information.
The code is available at \url{https://github.com/DakeZhang1998/ReadProbe}.
An earlier version of our tool won the first prize in a national AI misinformation hackathon.
\blfootnote{\copyright 2023 Copyright held by the authors.}
\end{abstract}

\ccsdesc[500]{Information systems~Decision support systems}

\keywords{Generative Information Retrieval, Lateral Reading, Large Language Models, Misinformation}



\settopmatter{printfolios=true}
\maketitle

\section{Introduction}

The Internet and social media networks have made the spread of information much faster and broader than ever before, but also open channels for the wild proliferation of misinformation.

Existing approaches to counter online misinformation can be classified into two broad categories: system-focused and individual-focused.
Examples of the former class are content moderation techniques used by many online platforms (like Twitter and Facebook), e.g., hiding posts identified as misinformation,  demoting low-quality information in their recommender systems to slow down the transmission of misinformation, flagging malicious posts, and so on.
However, even with good intentions, content moderation could raise concerns about limiting the freedom of expression~\cite{douek2021governing}, especially since those algorithms can not achieve high enough precision, leading to many false positives (posts misclassified as misinformation).

On the contrary, individual-focused approaches aim to improve the competency of readers (consumers of online information) to spot misleading information and avoid helping to spread it, such as psychological inoculation against misinformation through games~\cite{basol2020good}, prompting people to think twice before sharing posts~\cite{fazio2020pausing}, and developing curriculum for education~\cite{wineburg2022lateral}.
Lateral reading is also a promising approach of this kind, which was recently proposed~\cite{wineburg2019lateral}.
Different from traditional fact-checking methods that give authoritative judgments~\cite{zhang2022learning}, lateral reading aims to train readers to cross-reference multiple sources to better assess the correctness of online information.
Specifically, when consuming a web page, lateral readers frequently leave the original page and open new browser tabs to search for information from other sources.
Lateral reading has been shown to be an effective technique to evaluate online information~\cite{wineburg2019lateral, wineburg2022lateral}.

Existing studies on lateral reading focus on educational practices to enhance the media literacy of people.
To the best of our knowledge, there is no existing tool that can effectively support readers to perform lateral reading, such as coming up with questions for a web document and summarizing answers based on information from other sources.
Search engines are an essential component of lateral reading, but previous research has observed that many readers are not good at formulating queries or browsing search results to get the answers they want~\cite{wineburg2019lateral}.
Therefore, to bridge this gap, we developed \rp, as a demo of concept, to help people perform accurate knowledge-grounded lateral reading, which leveraged the reasoning capabilities of modern Large Language Models (LLMs) with augmentation of access to search engine results.

\section{Related Work}

\subsection{Lateral Reading}
\begin{figure*}
    \centering
    \includegraphics[width=0.9\linewidth]{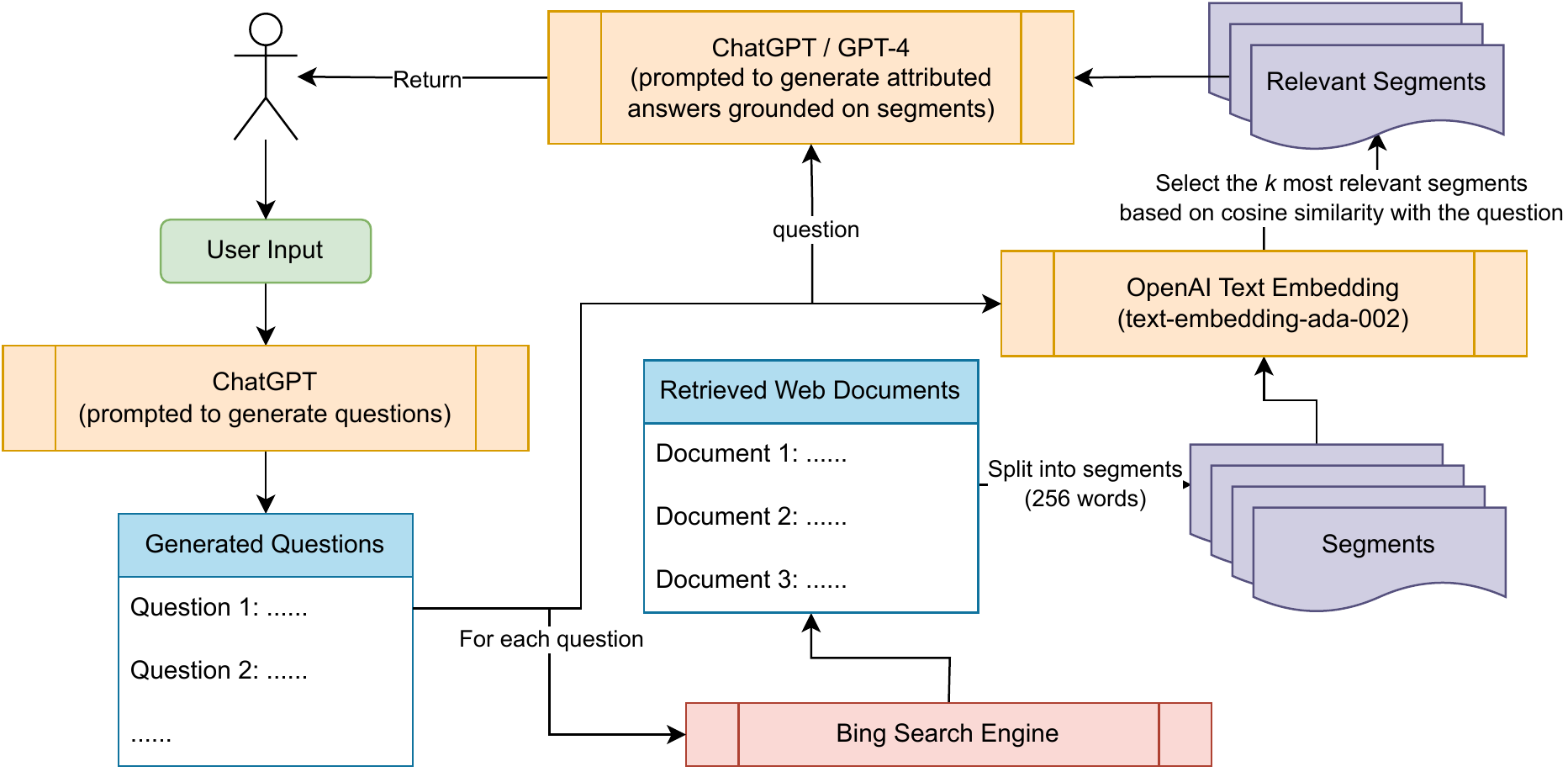}
    \caption{Illustration of Our \rp Pipeline.}
    \label{fig:overview}
\end{figure*}

\citet{wineburg2019lateral} first identified this strategy of lateral reading when observing how students, historians, and professional fact-checkers evaluated online information.
Different from students and historians, fact-checkers quickly left the original web pages and opened new tabs for searching, instead of sticking to the landing pages and reading through them.
It turned out that this observed technique made fact-checkers obtain more accurate assessments of the credibility of online information in a shorter time, compared with other people.

There have been a lot of ongoing studies on how to better educate people about lateral reading and evaluate their abilities to apply this strategy~\cite{mcgrew2020learning, brodsky2021improving, wineburg2022lateral, panizza2022lateral}.
So far, researchers have not reached a universally accepted definition of lateral reading.
In this paper, we take the core intuition of lateral reading that people should consume online information laterally (cross-referencing relevant information from multiple sources) instead of vertically (focusing on the original source and reading it from top to bottom).
Lateral reading has been collected as one of the effective individual-focused interventions against online misinformation in the survey from \citet{kozyreva2022toolbox}.
This survey was used to inspire participants in the 2023 Canadian \#AI Misinformation Hackathon~\cite{hackathon}.

\subsection{Large Language Models}

Recent advancements in Natural Language Processing (NLP) have given rise to the development of powerful generative LLMs such as GPT series~\cite{gpt3} and PaLM series~\cite{palm}.
These decoder-only models with over 100 billion parameters were pre-trained on vast amounts of textual data and demonstrated strong capabilities across various NLP tasks in zero-shot or few-shot settings.


Meanwhile, instruction fine-tuning of these pre-trained LLMs based on supervised learning~\cite{flanpalm} or reinforcement learning~\cite{instructgpt} from human feedback enabled them to adapt to a wide range of NLP tasks with human instructions (prompting). 
ChatGPT\footnote{\url{https://openai.com/blog/chatgpt}} is one such popular model fine-tuned with Reinforcement Learning from Human Feedback (RLHF), thereby adapting to the chat domain and following instructions in a prompt to provide detailed responses.
One drawback is that the model can only take as input up to 4096 tokens, which requires additional efforts in tasks that involve long documents.
Its successor, GPT-4~\cite{gpt4}, further built on this success and obtained the capability to reason across images and texts.
A variant of GPT-4 is capable of processing up to 32k tokens, thereby enabling reasoning across multiple long documents.

\section{Implementation Details}

Figure~\ref{fig:overview} shows an overview of our tool \rp.
It consists of three core components --- question generation, web search, and answer generation, each feeding into the next and coupled into an easy-to-use, tightly-knit application.
The user input can be a claim or a post consisting of one or several sentences, or a long web document in plaintext such as an online news article. 
\rp will generate several questions based on the provided text for users to probe into and generate answers by summarizing information from other sources.
In the rest part of this section, we discuss each of these components in detail.
To clarify, ChatGPT and GPT-4 in this section correspond to the models accessible via API calls to \texttt{gpt-3.5-turbo} and \texttt{gpt-4-32k}, respectively, as of May 1st, 2023.

\subsection{Question Generation}

The task of question generation from text has been at the core of Information Retrieval (IR) research, with applications such as serving as synthetic training data to address a lack of resources~\cite{0shotneural} and document augmentation for enhancing first-stage retrieval~\cite{d2q}.
In the case of \rp, we instead focus on the following question: how do we generate useful questions that, when viewed with well-attributed answers grounded on multiple sources from the Internet, will aid users to better understand and evaluate the credibility and accuracy of online information?

\begin{figure*}
\begin{mdframed}
\begin{verbatim}
Role: You are a factual and helpful assistant to aid users in the lateral reading task. You will receive a segment 
of text (Text:), and you need to raise five important, insightful, diverse, simple, factoid questions that may
arise to a user when reading the text but are not answered by the text (Question1:, Question2:, Question3:,  
Question4:, Question5:). The questions should be suitable as meaningful queries to a search engine like Bing.
Your questions will motivate users to search for relevant documents to better determine whether the given text
contains misinformation.
User: Text: {user_input}
Carefully choose insightful and atomic lateral reading questions not answered by the above text, ensuring that 
the queries are self-sufficient (Do not have pronouns or attributes relying on the text, they should be fully 
resolved and make complete sense independently).
\end{verbatim}
\end{mdframed}
\caption{Prompt for Question Generation.}
\label{fig:prompt_qgen}
\end{figure*}

Figure~\ref{fig:prompt_qgen} showcases the prompt we use to ask the GPT model to generate questions to support lateral reading.
It begins with a preamble that describes its role in tackling the task --- factuality and helpfulness at the core aiming to generate insightful and diverse questions that are meaningful to pass on to a search engine.
Next, the user input is formatted and provided to the model in the form of a chat response, followed by some closing information on how the queries need to reflect independent search queries and should not rely on coreference resolution conditioned on the user input.
In our tests, we found that explicitly mentioning this helps to generate higher-quality queries and also avoid being manipulated by malicious user inputs, such as DAN\footnote{\url{https://gist.github.com/coolaj86/6f4f7b30129b0251f61fa7baaa881516}}.

\subsection{Web Search and Segment Retrieval}

Web search is crucial to the success of \rp, as it helps overcome some shortcomings of LLMs regarding tail entities or lack of information available post the training corpora cut-off date.
In this web search component, we independently query each generated question using the Microsoft Bing web search API\footnote{\url{https://www.microsoft.com/en-us/bing/apis/bing-web-search-api}} and pass the top $n=3$ results on to the next stage (segment retrieval or answer generation, depending on which GPT model is used).
The hope is that such knowledge grounding will help prevent model hallucinations, reduce pre-conceived biases introduced during training, and provide users with resources that they can read further.

Our initial implementation of \rp was before the launch of the public-facing GPT-4 API, a variant of which can handle 32k tokens, enough to reason across multiple documents.
Hence, we tackled the issue of the input token limit (4k tokens) of ChatGPT by splitting the web page content into 256-word text segments and using the text similarity model from OpenAI to select the $k$-most similar segments to the query (based on their cosine similarity) from each retrieved web page ($k$ = 2 in \rp) as inputs for the answer generation.

Note that even with GPT-4's capability of handling long documents, we still believe segment retrieval to be quite useful for a few reasons.
First, the response time of GPT-4 is noticeably longer than ChatGPT, given it is bigger architecturally than ChatGPT, and possibly it is still in the testing phase.
Second, GPT-4 API usage costs an order of magnitude more than ChatGPT (especially for the variant that can handle 32k tokens), so we only want to feed the best segments to save costs.
Finally, based on our preliminary qualitative analysis of the generated answers from GPT-4, we did not observe any noticeable quality difference in the results with and without segment retrieval, except that the response is much cheaper and faster with segment retrieval.

\subsection{Answer Generation}

This stage contains the core downstream task, i.e.,  given a lateral reading question and relevant documents or text segments, to generate a well-attributed, useful, and  readable answer.
Proper attributions in the generated answers would help users build trust.

\begin{figure*}
\begin{mdframed}
\begin{verbatim}
Role: You are a factual and helpful assistant designed to read and cohesively summarize segments from different 
relevant document sources to answer the question at hand. Your answer should be informative but no more than 100
words. Your answer should be concise, easy to understand and should only use information from the provided 
relevant segments but combine the search results into a coherent answer. Do not repeat text and do not include 
irrelevant text in your answers. Use an unbiased and journalistic tone. Make sure the output is in plaintext. 
Attribute each sentence with proper citations using the document number with the [${doc_number}] notation 
(Example: "Hydroxychloroquine is not a cure for COVID-19 [1][3]."). Ensure each sentence in the answer is 
properly attributed. Ensure each of the documents is cited at least once. If different results refer to different 
entities with the same name, cite them separately.
User: My question is {question}. Cohesively and factually summarize the following documents to answer my 
question. {doc_texts}
\end{verbatim}
\end{mdframed}
\caption{Prompt for Answer Generation.}
\label{fig:prompt_answer}
\end{figure*}

Figure~\ref{fig:prompt_answer} shows the prompt we use to instruct the GPT model to generate answers.
This prompt was inspired by other successful retrieval-in-the-loop systems like perplexity\footnote{\url{https://www.perplexity.ai/}}.
The preamble or role description of the agent describes the need to be factual while cohesively stringing together information across sources into a short but informative answer. 
The prompt promotes an unbiased and journalistic tone and dissuades the use of extraneous information in the answer.
Then we ensure that attribution is at the core by forming a particular attribution schema and providing an example sentence.
Note that while we specified to cover all the sources, we still found cases where a relevant web page was not used at all in the generated answer.
In such a case, we will mention it when rendering results.

Finally, as a chat input from a user, the lateral reading question and some words reiterating the task details are fed into the GPT model, followed by the relevant segments.
The only hyperparameter we tune is \texttt{temperature}, and we find that a value of $0.2$ works as a nice balance between diversity and quality.

\section{System Demonstration}

We developed a web-based application to demonstrate \rp built with \texttt{Streamlit}\footnote{\url{https://streamlit.io/}}.
The code is available at \url{https://github.com/DakeZhang1998/ReadProbe}.

\subsection{Interface}

Figure~\ref{fig:demo-screenshot-main} in Appendix~\ref{sec:appendix} shows the main interface of \rp.
We designed this interface to provide sufficient background information and clear instructions for users who do not know the concept and benefits of lateral reading beforehand.
On the sidebar of the main interface, we briefly describe what lateral reading is, what purpose it serves in the fight against online misinformation, and how our tool can support lateral reading.
Users can provide text in the input box, either several sentences or a document (with a limit of 2000 words).
Then all they need to do is to click on the ``Probe'' button, and our system will start working to generate questions and answers to support lateral reading.
Our privacy policy, which we noticeably place on the landing page, states that OpenAI and we may collect some anonymized data.
An instance on the OpenAI end could be that the user input text, which we embed with our OpenAI API requests, may be leveraged in improving their models.
By using \rp, we hope that in its current iteration, users are aware when consenting to us and OpenAI potentially leveraging their anonymous data and feedback.
In future iterations, as \rp becomes stable and its constituent OpenAI API calls can guarantee not collecting or training with user data, we hope to provide the same choice to users.

\subsection{Opinion Probing - A Qualitative Example}

To demonstrate how \rp works, we provide a demo input, a multi-sentence opinion against the use of vaccinations for COVID-19 and beyond, seen in Figure~\ref{fig:demo-screenshot-main} in Appendix~\ref{sec:appendix}.
This opinion goes into how vaccines insert microchips and calls for a stand against future cancer and heart disease vaccines.

Once the user clicks on the ``Probe'' button, the system will return a list of questions with attributed answers from three sources, shown in Figure~\ref{fig:demo-screenshot-questions} in Appendix~\ref{sec:appendix}.
If the user likes a specific question and its answer, they can click on the ``I like this one'' button to indicate their preferences.

Given this input, we first see that \rp provides a good set of diverse questions covering various points made in the opinion.
The first two questions help the user understand that there is a lack of evidence for the supposed link between vaccines and youth death rates from a governmental and scientific perspective.
The third question-answer pair helps debunk the microchip claims.
The next pair helps the user understand mRNA vaccines, and the final looks at the future of vaccines, for cancer, heart disease, and beyond.
We see that this covers most ideas originating from the example input and hope that such information from \rp will help decision-makers and users make informed choices. 

\subsection{Anonymous User Feedback}

We build a lightweight online database with Google Forms to collect anonymous user feedback.
Specifically, we collect only their input text and their feedback (clicks on the ``I like this one'' button) so that we can evaluate our system and improve it.
We collect this data strictly anonymously, without any fields that one could use to identify users, such as session identifiers, IP addresses, and locations.
As mentioned, we disclose the information we collect in our privacy policy on the main interface.

\section{Conclusion}

In this paper, we present our LLM-powered tool \rp, designed to support lateral reading to help build people's competency to fight against online misinformation.
To the best of our knowledge, this is the first attempt to develop tools to aid the lateral reading strategy in this age of LLMs.
\rp decomposes the problem and utilizes various components, including question generation, web search, and answer generation, to assist users in evaluating the credibility and accuracy of online information while avoiding model hallucinations with knowledge grounding. We build \rp with careful consideration of issues stemming from input token limit, response time, and cost-effectiveness. Overall, the tool shows promise in helping users navigate the vast and sometimes confusing landscape of online information.
We hope our attempt will inspire other academic and industry practitioners and open up new lines of work.

\textbf{Future Work.} In our pilot user study, this tool garnered overwhelmingly positive feedback on its usefulness for identifying misinformation.
We also improved our tool according to suggestions from users.
However, a large-scale user study (with a control group and a treatment group) is needed to demonstrate the usefulness of \rp for evaluating online information.
Additionally, since there are a lot of countries in the world where misinformation proliferates in local languages, it remains to be seen how \rp would work when extended to a multilingual setup.
However, given the additional complexities, we leave this important extension for work to come.

\begin{acks}
This work was supported in part by Microsoft, in part by the Natural Sciences and Engineering Research Council of Canada (RGPIN-2020-04665, RGPAS2020-00080), and in part by the University of Waterloo.
We would also like to thank members of the Social Media Lab at the Toronto Metropolitan University for organizing the 2023 Canadian \#AI Misinformation Hackathon.
\end{acks}

\bibliographystyle{ACM-Reference-Format}
\bibliography{mypaper}


\appendix

\section{Application Demonstration} \label{sec:appendix}

See the next page for two screenshots taken from our web application.

\begin{figure*}[h]
    \centering
    \includegraphics[width=0.9\linewidth]{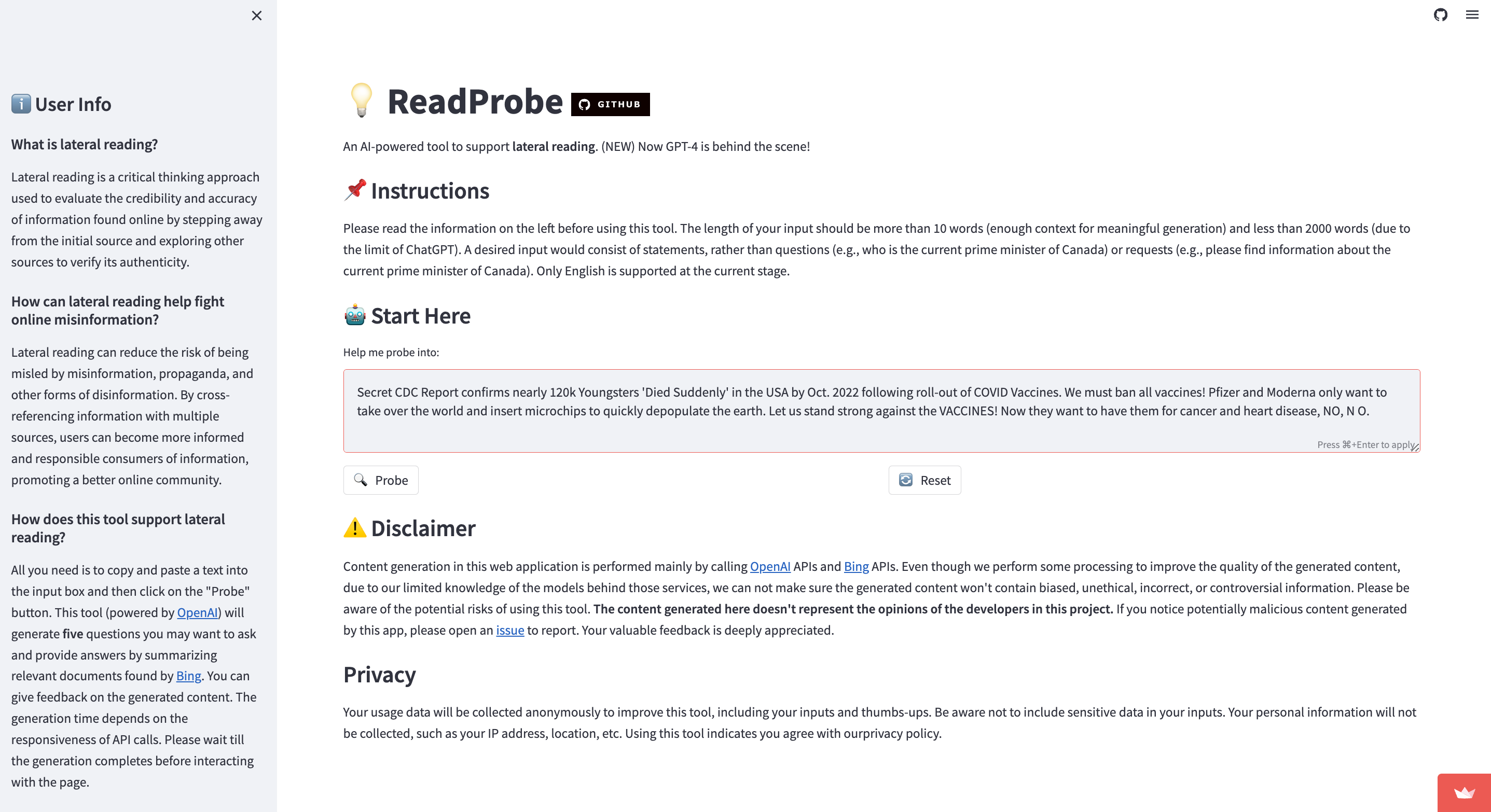}
    \caption{Main Interface of \rp.}
    \label{fig:demo-screenshot-main}
\end{figure*}

\begin{figure*}[h]
    \centering
    \includegraphics[width=\linewidth]{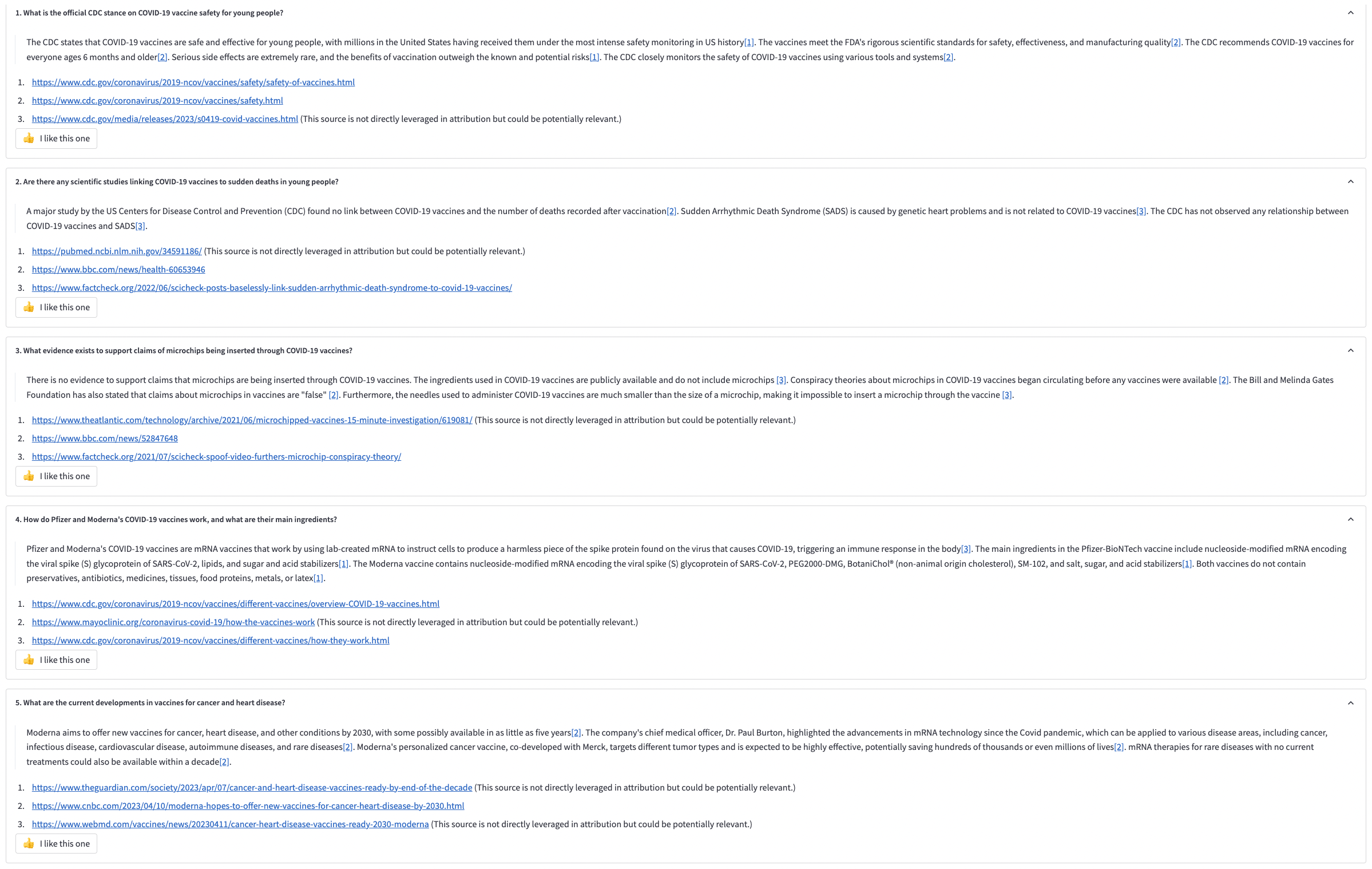}
    \caption{Top Five Questions with Answers Generated for the Demo Claim.}
    \label{fig:demo-screenshot-questions}
\end{figure*}

\end{document}